\documentclass[10pt]{article}
\usepackage{graphicx}
\usepackage{longtable}
\title{On magnetic fields of radio pulsars}

\author{
Nikitina E., Malov I.\\
Pushchino Radioastronomical Observatory, 142290~Moscow Region, Pushchino, Russia}

\begin{document}

\maketitle
\begin{abstract}
We used the magneto-dipole radiation mechanism for the braking of radio pulsars to calculate the new values of magnetic inductions at the surfaces of neutron stars. For this aim we estimated the angles $\beta$ between the rotation axis and the magnetic moment of the neutron star for 376 radio pulsars using three different methods. It was shown that there was the predominance of small inclinations of the magnetic axes. Using the obtained values of the angle $\beta$ we calculated  the equatorial magnetic inductions for pulsars considered. These inductions are several times higher as a rule than  corresponding values in the known catalogs.

{\bf keywords}
magnetic fields; methods: data analysis; methods: statistical; \textit{(stars:)} pulsars: general
\end{abstract}

\section{Introduction}

It is common accepted that braking of pulsars is caused by the magneto-dipole radiation of the rotating magnetic star. In this case the rate of losses of the neutron star rotation energy can be equated to the power of its magneto-dipole radiation:
\vspace{\baselineskip}
\begin{equation}
-I\Omega \frac{d\Omega}{dt}=\frac{2\Omega^4\mu^2\sin^2\beta}{3c^3},
\end{equation}
\\where \textit{I} is the moment of inertia of the neutron star, $\Omega$ - the angular speed of its rotation, $\mu$ - its magnetic moment, $\beta$ - the angle between the rotation axis and the magnetic moment, \textit{c} - speed of light. For standard parameters of neutron stars: masses of order of the solar mass ($2\times 10^{33}g$) and radii \textit{R} of order of $10^6$ cm we can put \textit{I} = $10^{45}g\times cm^2$. For the magnetic moment we have
\begin{equation}
\mu=\frac{B_PR^3}{2}=B_{eq}R^3.
\end{equation}
\\Here $B_P$ is the magnetic induction at the magnetic pole, $B_{eq}$ ? the induction at the magnetic equator. Instead of $\Omega$ the rotation period $P = 2\pi/\Omega$ is usually measured and we can obtain from (1) and (2):
\vspace{\baselineskip}
\begin{equation}
B_{P}^2\sin^2\beta=\frac{3Ic^3P\dot{P}}{2\pi^2R^6}=4.1\times10^{39}P\dot{P}.
\end{equation}
\\This equality is used usually to calculate magnetic inductions of pulsars assuming that $\sin\beta=1$ for all objects. The known catalogs (see, for example Manchester et al., 2005) contain as a rule $B_{eq}$ instead of $B_P$. Here we propose to decline the assumption on the constancy of $\sin\beta$ and use some estimations of this parameter to calculate more accurate values of pulsar magnetic inductions.

\section{Methods of estimations of the angles between magnetic and rotation axes}
In a number of our works (Malov \& Nikitina, 2011a,b, 2013) some methods for calculations of the angle $\beta$ have been put forward and applied to some catalogs of pulsars (Keith et al., 2010; van Ommen et al., 1997; Weltevrede \& Johnston, 2008) at approximately 10, 20 and 30 cm. Basic equations for this aim are (Manchester \& Taylor, 1977):
\vspace{\baselineskip}
\begin{equation}
	\cos\theta=\cos\beta\cos\zeta+\sin\zeta\sin\beta\cos\Phi_P,
\end{equation}

\begin{equation}
  \tan\psi=\frac{\sin\beta\sin\Phi}{\sin\zeta\cos\beta-\cos\zeta\sin\beta\cos\Phi}.
\end{equation}
\\Here $\zeta$ is the angle between the line of sight and the rotation axis, $\theta$ - the angular radius of the emission cone, $\Phi_P$ - a half of the angular width of the observed pulse, $\psi$ - the position angle of the linear polarization, $\Phi$ - longitude.

\subsection{Suggestion on the equality of angles $\zeta$ and $\beta$}
The simplest case for the calculations of the angle $\beta$ is realized when the line of sight passes through the center of the emission cone, i.e.
\begin{equation}
	\zeta=\beta.
\end{equation}
\\In this case we can use the dependence of the observed pulse width $W_{10}$ at the $10\%$ level on the rotation period and determine the lower boundary in the corresponding diagram to obtain
\vspace{\baselineskip}
\begin{equation}
	\theta=\frac{W_{10min}}{2}.
\end{equation}
\\As the result we have from (4), (5) and (7) (Malov \& Nikitina, 2011a):

\begin{equation}
	\sin\beta=\frac{\sin\frac{W_{10min}}{4}}{\sin\frac{W_{10}}{4}}.
\end{equation}
\\The values of angles calculated by this method are denoted as $\beta_1$ and given in the Table 1.

\subsection{Taking into account polarization data}
Usually polarization measurements are made inside the pulse longitudes only. In this case we can use the maximal derivative of the position angle. From (5) we have
\vspace{\baselineskip}
\begin{equation}
C=\left(\frac{d\psi}{d\Phi}\right)_{max}=\frac{\sin\beta}{\sin\left(\zeta-\beta\right)}.	
\end{equation}
\vspace{\baselineskip}

We can obtain from the dependence of $W_{10}$ on \textit{P} by the least squares method
\vspace{\baselineskip}
\begin{equation}
\theta=\frac{\left\langle W_{10}\right\rangle}{2}.		
\end{equation}
\\The third equation for the calculations of the angle $\beta$ is (4). From these three equations we obtain
\vspace{\baselineskip}
\begin{equation}
C^2(1-D)^2y^4+2C(1-D)y^3+\left[1+2C^2D(1-D)\right]y^2+2C(D-B^2)y+C^2D^2-B^2(1+C^2)=0.\\
\end{equation}
\\Here

\begin{equation}
B=\cos\theta, D=\cos\frac{W_{10}}{2}, y=\cos\zeta.			
\end{equation}
\\We can transform the equation (9) to the following form

\begin{equation}
\tan\beta=\frac{C(1-y^2)^{\frac{1}{2}}}{1+Cy}.			
\end{equation}
\\Then finding the value of y from the equation (11) we can calculate $\beta$ from (13).

We have calculated values of $\beta$ by this method and list them in the Table 1 as $\beta_2$. Here we correct the misprint in the equation (11) made in our papers (Malov \& Nikitina, 2011a,b, 2013).

\subsection{Using position angles and shapes of average profiles}
There is an additional way to calculate angles $\beta$. This way uses observable values of position angles and shapes of average profiles for individual pulsars. In this case, original equations form the closed system for calculations of the angles $\theta$, $\zeta$ and $\beta$:

\begin{equation}
\begin{array}{rcl}
\sin \beta & = & C\sin (\zeta - \beta),\\
\cos \theta & = & \cos \zeta \cos \beta + D \sin \beta \sin \zeta,\\
\theta & = & n (\zeta - \beta),
\end{array}
\end{equation}

\vspace{\baselineskip}
As the observed pulsar profiles have various forms, the coefficient \textit{n} has a different value depending on a profile structure. We put arbitrary the following values of \textit{n} (Fig.1). If the ratio of the intensity $I_0$ in the center of the pulse to the maximal intensity $I_{max}$ is zero then $n = \infty$. For $I_0/I_{max}<1/2$ $n=4$, $I_0/I_{max} = 1/2$ $n=2$, $I_0/I_{max}>1/2$ $n=3/2$, and for $I_0/I_{max} \approx 1$ $n=5/4$. It is worth noting that the solution of the system (14) can be obtained numerically for any value of \textit{n}.

\begin{figure}
\begin{center}
		\includegraphics[width=12cm]{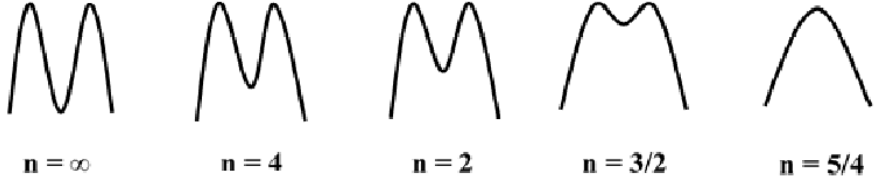}
    \caption{Different types of profiles.}
    \label{fig:Fig1.png}
		\end{center}
\end{figure}

For example, if $n = 4$, the solution for $y =\cos\zeta$ can be obtained from the equation:

\begin{equation}
(C D + y + C y^2 (1 - D)) \sqrt{(C^2 + 2 C y + 1)^3} - 8 y^4 - 16 C y^3-4(3 C^2 - 2) y^2 - 4C(C^2 - 3) y - C^4 + 6 C^2 - 1 = 0;
\end{equation}

\vspace{\baselineskip}
at n = 2:

\begin{equation}
\begin{array}{lcr}
2 C^3 (1 - D)^2 y^5 + \left[ C^4 (1 - D)^2 + C^2 (D^2 - 6D + 5) - 4 \right] y^4 + 2C \left[ C^2 (1 + D - 2D^2) - 2 - D \right] y^3 + \left[ 2DC^4 (1 - D) - \right. \\
\left. - C^2 (2D^2 - 6D + 7) + 5 \right] y^2  + 2C \left[ C^2 D^2 + D(1 + C^2) - 2 (C^2 - 1) \right] y + C^2 D^2 (1 + C^2) - (C^2 - 1)^2 = 0;\\
\end{array}
\end{equation}

\vspace{\baselineskip}
at n = 3/2:

\begin{equation}
\left[2 (y + C) - \sqrt{C^2 + 2Cy + 1}\right] \sqrt{\frac{1 + \frac{C + y}{\sqrt{C^2 + 2Cy +1}}}{2}} - C y^2 (1 - D) - y - CD = 0;
\end{equation}

\vspace{\baselineskip}
at n = 5/4:

\begin{equation}
\left( 1 + \frac{2(C + y)}{\sqrt{C^2 + 2Cy +1}} - \sqrt{2 \left( 1
+ \frac{C + y}{\sqrt{C^2 + 2Cy +1}} \right)} \right) \sqrt{\frac{1 + \sqrt{\frac{1 + \frac{C + y}{\sqrt{C^2 + 2Cy +1}}}{2}}}{2}} - \frac{C y^2 (1 - D) + y + CD}{\sqrt{C^2 + 2Cy +1}} = 0.
\end{equation}

\vspace{\baselineskip}
This method gives angles $\beta_3$ (see the Table 1).

For some pulsars calculations were made by one method only. When it was possible we used two or all three methods. In these cases, the mean value of the angle $\beta$ has been calculated. The resulting values $\left\langle \beta\right\rangle$ are listed in the Table 1. Some other authors (for example, Kuz'min \& Dagkesamanskaya, 1983; Kuz'min et al., 1984; Lyne \& Manchester, 1988) carried out calculations of the angle $\beta$ earlier for the shorter samples of pulsars using some additional assumptions.

We will use further our estimations to calculate magnetic inductions at the surface of the neutron stars. The distribution of the angles $\beta$ from the Table 1 (Fig.2) shows that the majority of pulsars have rather small inclinations of the magnetic moments. These pulsars are old enough, and we can conclude that they evolve to the aligned geometry. The average characteristic age for our sample of pulsars is $7.96\times 10^7$ years. We must note however that the angles calculated by the method \textbf{\textit{1)}} are the lower limits of this parameter. This explains partly the predominance of the small values of $\beta$.

\begin{figure}
		\includegraphics[width=12cm]{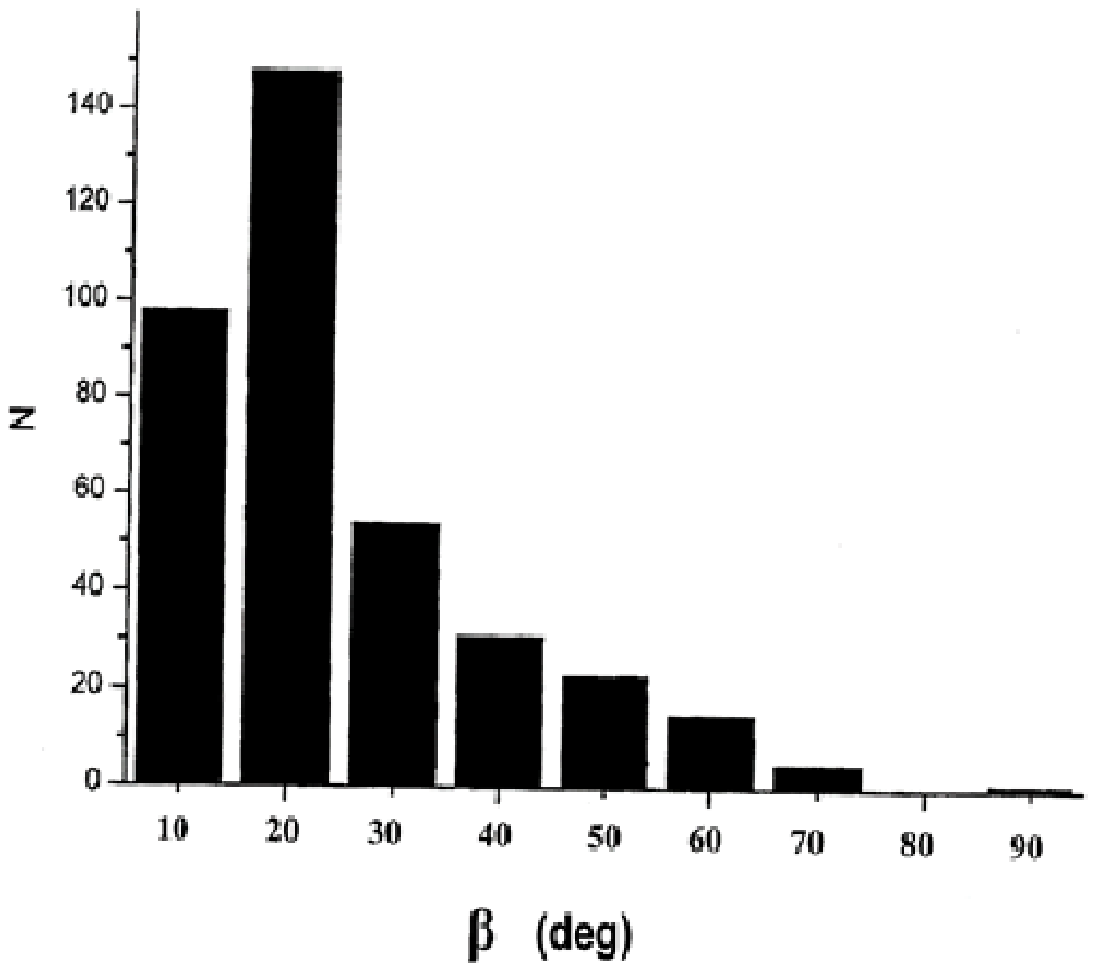}
    \caption{Distribution of mean values of the angle $\beta$ from the Table 1.}
\end{figure}

\vspace{\baselineskip}
\begin{longtable}{|c|c|c|c|c|c|c|c|c|c|c|}
\caption[Values of the angle $\beta$ (deg).]{Values of the angle $\beta$ (deg).}
\\
\hline
	 & $PSR$ & $\beta_1$ & $\beta_1$ & $\beta_1$ & $\beta_2$ & $\beta_2$ & $\beta_2$& \multicolumn{2}{c|}{$\beta_3 30 cm$} & $\left\langle\beta\right\rangle (deg)$\\
\cline{9-10}
 & &10 cm&20 cm&30 cm&10 cm&20 cm&30 cm&$C>0$&$C<0$& \\
\hline
\endhead
\hline
\endfoot
1&J0034-0721& &5&9& & & & & &7\\
2&J0051+0423&	&5&	&	&	&	&	&	&5\\
3&J0108-1431&8&7&	&11&11&	&	&	&9\\
4&J0134-2937& &16& & & & & & &16\\
5&J0151-0635&	&4& &	&	&	&	&	&4\\
6&J0152-1637&	&20&33&	&	&12&64&	&32\\
7&J0206-4028&	&20&40&	&	&	&	&46&35\\
8&J0211-8159&	&5&	&	&	&	&	&	&5\\
9&J0255-5304&	&26&44&	&	&	&	&	&35\\
10&J0304+1932& &11&	&	&	&	&	&	&11\\
11&J0401-7608&12&12&19&48& & & & &23\\
12&J0448-2749& &13&	&	&	&	&	&	&13\\
13&J0450-1248& &7& & & & & & &7\\
14&J0452-1759&15&	&	&	&	&	&	& &15\\
15&J0459-0210& &13&15& & & & & &14\\
16&J0520-2553& &13& & & & & & &13\\
17&J0525+1115& &13& & & & & & &13\\
18&J0533+0402& &18& & & & & & &18\\
19&J0536-7543&7&6&11&28&33&27& &50&23\\
20&J0540-7125& &11& & & & & & &11\\
21&J0543+2329& &11& & & & & & &11\\
22&J0601-0527& &10& & & & & & &10\\
23&J0614+2229&17&18& &44& & & & &26\\
24&J0624-0424& &8& & & & & & &8\\
25&J0630-2834&6&5&8&18&21&16& &65&20\\
26&J0631+1036& &10& & & & & & &10\\
27&J0636-4549& &27& & & & & & &27\\
28&J0656-2228& &20& & & & & & &20\\
29&J0656-5449& &14& & & & & & &14\\
30&J0659+1414&8&8& &17&21& & & &13\\
31&J0709-5923& &30& & & & & & &30\\
32&J0719-2545& &16& & & & & & &16\\
33&J0729-1448&10&9& & & & & & &10\\
34&J0729-1836& &12& & & & & & &12\\
35&J0738-4042& & &11& & &21& & &16\\
36&J0742-2822&19&18&32&37&42&49& &24&32\\
37&J0745-5353&9&8&16& &26&23&53&54&27\\
38&J0749-4247& &20& & & & & & &20\\
39&J0809-4753& & &24& & & & & &24\\
40&J0820-1350& & &29& & & &76&71&59\\
41&J0820-3921& &4& & & & & & &4\\
42&J0821-4221& &10& & & & & & &10\\
43&J0834-4159&16&13& & & & & & &15\\
44&J0837+0610& & &32& & & & & &32\\
45&J0837-4135& & &49& & &58& & &54\\
46&J0838-2621& &7& & & & & & &7\\
47&J0843-5022& &14& & & & & & &14\\
48&J0846-3533& & &12& & & & & &12\\
49&J0849-6322& &10& & & & & & &10\\
50&J0855-3331& & &36& & & &59&47&47\\
51&J0856-6137& &13& & & & & & &13\\
52&J0857-4424&14&13& & & & & & &14\\
53&J0901-4624&13&8& & & & & & &11\\
54&J0902-6325& &12& & & & & & &12\\
55&J0905-4536& &2& & & & & & &2\\
56&J0905-5127&18&19& & &22& & & &20\\
57&J0907-5157& &8&13& & &13&54& &22\\
58&J0908-4913&32&28& &72& & & & &44\\
59&J0922+0638& & &30& & &38&66&32&42\\
60&J0924-5302& &18& & & & & & &18\\
61&J0924-5814& &5& & & & & & &5\\
62&J0932-3217& &30& & & & & & &30\\
63&J0934-4154& &13& & & & & & &13\\
64&J0934-5249& & &25& & &40& &55&40\\
65&J0941-5244& &15& & & & & & &15\\
66&J0942-5552& & &12& & &18&59&58&37\\
67&J0942-5657& &34& & & & & & &34\\
68&J0953+0755& & &15& & & & & &15\\
69&J0954-5430&23&16& & & & & & &20\\
70&J0955-5304& & &25& & & & & &25\\
71&J1001-5507& & &47& & & & & &47\\
72&J1003-4747&13&11& & & & & & &12\\
73&J1015-5719&3&2& &5&10& & & &5\\
74&J1016-5345& &29& & & & & & &29\\
75&J1016-5857& &8& & & & & & &8\\
76&J1017-5621& &22& & & & & & &22\\
77&J1019-5749&10& & &5& & & & &7\\
78&J1020-6026&6&3& & & & & & &5\\
79&J1032-5911& &8& & & & & & &8\\
80&J1034-3224& &2& & & & & & &2\\
81&J1036-4926& &18& & & & & & &18\\
82&J1038-5831& &17& & & & & & &17\\
83&J1043-6116&19&19& & & & & & &19\\
84&J1046-5813& &19& & & & & & &19\\
85&J1047-3032& &5& & & & & & &5\\
86&J1047-6709& &14& & & & & & &14\\
87&J1048-5832&15&11& &69&73& & & &42\\
88&J1052-5954&26&14& & & & & & &20\\
89&J1055-6028&14&10& & & & & & &12\\
90&J1056-6258& & &12& & &17&46&16&23\\
91&J1057-5226&8&8&18&15&24&18& & &15\\
92&J1059-5742& & &32& & &55&74& &54\\
93&J1110-5637& &12& & & & & & &12\\
94&J1112-6613& &11& & & & & & &11\\
95&J1112-6926& &12& & & & & & &12\\
96&J1114-6100& &9& & & & & & &9\\
97&J1115-6052&22&23& & &35& & & &27\\
98&J1116-4122& & &46& & & & & &46\\
99&J1119-6127&7&5& &17&18& & & &12\\
100&J1123-4844& &13& & & & & & &13\\
101&J1123-6259&19&15& & &43& & & &26\\
102&J1126-6054& &19& & & & & & &19\\
103&J1126-6942& &16& & & & & & &16\\
104&J1133-6250& &2& & & & & & &2\\
105&J1136+1551& & &29& & &55& &45&43\\
106&J1137-6700& &2& & & & & & &2\\
107&J1138-6207&30& & & & & & & &30\\
108&J1141-3107& &11& & & & & & &11\\
109&J1141-3322& &9& & & & & & &9\\
110&J1143-5158& &22& & & & & & &22\\
111&J1146-6030& &14& & & & & & &14\\
112&J1157-6224& & &19& & & & & &19\\
113&J1204-6843& &18& & & & & & &18\\
114&J1215-5328& &7& & & & & & &7\\
115&J1216-6223&9&11& & & & & & &10\\
116&J1224-6407&21&19&38& & & & &27&26\\
117&J1225-5556& &17& & & & & & &17\\
118&J1225-6408& &12& & & & & & &12\\
119&J1231-4609& &7& & & & & & &7\\
120&J1236-5033& &9& & & & & & &9\\
121&J1240-4124& &31& & & & & & &31\\
122&J1243-6423& & &39& & & & &78&59\\
123&J1244-5053& &18& & & & & & &18\\
124&J1248-6344& &6& & & & & & &6\\
125&J1253-5820& &15& & & & & & &15\\
126&J1301-6305&4&3& & & & & & &4\\
127&J1305-6203&12&10& & &35& & & &19\\
128&J1305-6455& &6& & & & & & &6\\
129&J1306-6617& &5& & & & & & &5\\
130&J1319-6056& &17& & & & & & &17\\
131&J1320-3512& &10& & & & & & &10\\
132&J1320-5359&13&13& & &61& & & &29\\
133&J1326-5859& & &16& & &24& &63&35\\
134&J1327-6222& & &20& & & & & &20\\
135&J1327-6301& &6& & & & & & &6\\
136&J1327-6400&13&7& & & & & & &10\\
137&J1333-4449& &24& & & & & & &24\\
138&J1339-4712& &35& & & & & & &35\\
139&J1340-6456& &10& & & & & & &10\\
140&J1341-6220&21& & &73& & & & &47\\
141&J1349-6130&25&19& & &64& & & &36\\
142&J1352-6803& &9& & & & & & &9\\
143&J1356-5521& &9& & & & & & &9\\
144&J1357-6429&5&4& & &14& & & &8\\
145&J1359-6038&23&22&38& & &21&83& &37\\
146&J1401-6357& & &50& & & & &20&35\\
147&J1403-7646& &4& & & & & & &4\\
148&J1406-6121&43& & & & & & & &43\\
149&J1410-7404& &88& & & & & & &88\\
150&J1412-6145&18&11& & & & & & &14\\
151&J1413-6141&44& & & & & & & &44\\
152&J1413-6307& &33& & & & & & &33\\
153&J1415-6621& &18& & & & & & &18\\
154&J1427-4158& &14& & & & & & &14\\
155&J1428-5530& & &24& & &37& &33&31\\
156&J1430-6623& &25& & & & & & &25\\
157&J1452-5851&13&13& & &34& & & &20\\
158&J1453-6413&15&17&37& & &63& &70&40\\
159&J1456-6843&8&7&13&24& &25& &61&23\\
160&J1507-4352& &24& & & & & & &24\\
161&J1507-6640& &43& & & & & & &43\\
162&J1512-5759&14&10& & &65& & & &30\\
163&J1513-5908&3&4& & &10& & & &6\\
164&J1514-4834& &21& & & & & & &21\\
165&J1514-5925& & & &10&17& & & &14\\
166&J1515-5720& &13& & & & & & &13\\
167&J1517-4356& &18& & & & & & &18\\
168&J1522-5829& &9& & & & & & &9\\
169&J1524-5706&11&19& & & & & & &15\\
170&J1528-4109& &21& & & & & & &21\\
171&J1530-5327&18&15& & & & & & &17\\
172&J1531-4012& &16& & & & & & &16\\
173&J1534-5334& & &19& & & & & &19\\
174&J1534-5405& &10& & & & & & &10\\
175&J1535-4114& &12& & & & & & &12\\
176&J1536-3602& &6& & & & & & &6\\
177&J1538-5551&68& & & & & & & &68\\
178&J1539-5626&12&11& & &38& & & &20\\
179&J1541-5535&34&17& & & & & & &26\\
180&J1542-5034& &33& & & & & & &33\\
181&J1543-5459&46& & & & & & & &46\\
182&J1548-5607&10&9& & & & & & &9\\
183&J1549-4848&16&18& & & & & & &17\\
184&J1551-5310&19& & & & & & & &19\\
185&J1557-4258& &13& & & & & & &13\\
186&J1559-4438& & &25& & & & & &25\\
187&J1600-5044&33& &14& & &20& &22&22\\
188&J1600-5751&5&6& & & & & & &5\\
189&J1601-5335& &13& & & & & & &13\\
190&J1602-5100&15&17&29&65& & & & &32\\
191&J1603-3539& &8& & & & & & &8\\
192&J1603-5657& &54& & & & & & &54\\
193&J1604-4909& & &31& & & & & &31\\
194&J1605-5257& &5& & & & & & &5\\
195&J1607-0032& & &26& & & & & &26\\
196&J1609-1930& &28& & & & & & &28\\
197&J1611-5209&59&60& & & & & & &60\\
198&J1612-2408& &15& & & & & & &15\\
199&J1614-3937& &11& & & & & & &11\\
200&J1614-5048&18& & &70& & & & &44\\
201&J1615-5537& &18& & & & & & &18\\
202&J1626-4807&8& & & & & & & &8\\
203&J1632-4757&16& & & & & & & &16\\
204&J1632-4818&17& & & & & & & &17\\
205&J1633-5015& &14& & & & & & &14\\
206&J1637-4553&19&17& & & & & & &18\\
207&J1637-4642&6&5& &17&20& & & &12\\
208&J1638-4417&12&11& & & & & & &12\\
209&J1638-4608&66& & & & & & & &66\\
210&J1638-4725&9& & & & & & & &9\\
211&J1639-4604& &10& & & & & & &10\\
212&J1640-4715&15& & & & & & & &15\\
213&J1641-2347& &7& & & & & & &7\\
214&J1643-4505&18&12& & & & & & &15\\
215&J1644-4559& & &9& & &4& & &7\\
216&J1645-0317& & &14& & &1& & &8\\
217&J1646-4346&17& & & & & & & &17\\
218&J1646-6831& & &54& & & & & &54\\
219&J1648-4611&16&5& &26& & & & &16\\
220&J1649-4653&20& & & & & & & &20\\
221&J1649-5553& &3& & & & & & &3\\
222&J1650-1654& &12& & & & & & &12\\
223&J1650-4502&48& & & & & & & & 48\\
224&J1650-4921&40&27& & & & & & &33\\
225&J1651-4246& & &5& & &12& & &9\\
226&J1651-7642& &7& & & & & & &7\\
227&J1652-1400& &10& & & & & & &10\\
228&J1653-3838& &16& & & & & & &16\\
229&J1654-2713& &16& & & & & & &16\\
230&J1655-3048& &3& & & & & & &3\\
231&J1700-3312& &12& & & & & & &12\\
232&J1701-3726& &9& & & & & & &9\\
233&J1701-4533& &7& & & & & & &7\\
234&J1702-4128&11& & &32& & & & &21\\
235&J1702-4306& &11& & & & & & &11\\
236&J1702-4310&9&8& &11&15& & & &11\\
237&J1703-3241& & &21& & &62& &76&53\\
238&J1703-4851& &10& & & & & & &10\\
239&J1705-1906&16&14&22&49&70&45& &71&41\\
240&J1705-3950&10&7& &37&35& & & &22\\
241&J1709-1640& & &29& & &68& &77&58\\
242&J1709-4429&9&7& &20&28& & & &16\\
243&J1713-3949&59&20& & & & & & &39\\
244&J1714-1054& &27& & & & & & &27\\
245&J1715-3903&11&9& & &15& & & &12\\
246&J1717-5800& &6& & & & & & &6\\
247&J1718-3718&18& & & & & & & &18\\
248&J1719-4006& &14& & & & & & &14\\
249&J1721-3532&11& & &37& & & & &24\\
250&J1722-3207& & &29& & & & & &29\\
251&J1722-3632& &7& & & & & & &7\\
252&J1722-3712&25&20& &50&50& & & &37\\
253&J1723-3659&11&9& &28&33& & & &20\\
254&J1726-3530&40& & & & & & & &40\\
255&J1730-3350&27& & &61& & & & &44\\
256&J1731-4744&19&18&27& & & & &50&28\\
257&J1733-2228& &5& & & & & & &5\\
258&J1733-3716&5&4& &18&20& & & &12\\
259&J1734-3333&6& & &21& & & & &13\\
260&J1735-3258&10& & & & & & & &10\\
261&J1737-3137&18&7& & &20& & & &15\\
262&J1737-3555& &19& & & & & & &19\\
263&J1738-2955&19&22& & & & & & &21\\
264&J1739+0612& &10& & & & & & &10\\
265&J1739-1313& &50& & & & & & &50\\
266&J1739-2903&26&21& & & & & & &23\\
267&J1739-3023&16&16& & &36& & & &23\\
268&J1740-3015&44&39& & & & & & &42\\
269&J1741-3927& & &29& & &66& &72&56\\
270&J1742-4616& &6& & & & & & &6\\
271&J1743-3150& &13& & & & & & &13\\
272&J1743-3153& &7& & & & & & &7\\
273&J1743-4212& &13& & & & & & &13\\
274&J1745-3040&10&11&35& & & & & &19\\
275&J1749-3002& &4& & & & & & &4\\
276&J1750-3157& &6& & & & & & &6\\
277&J1751-4657& & &34& & & & & &34\\
278&J1752-2806& & &41& & & & & &41\\
279&J1755-2534& &6& & & & & & &6\\
280&J1756-2225&13&11& & & & & & &12\\
281&J1757-2421&9&9& & & & & & &9\\
282&J1759-2302& &4& & & & & & &4\\
283&J1801-2154&17&15& & &40& & & &24\\
284&J1801-2304&11& & & & & & & &11\\
285&J1801-2451&16&13& & & & & & &15\\
286&J1801-2920& &9& & & & & & &9\\
287&J1803-2137&3&3& &12&13& & & &8\\
288&J1803-2712& &7& & & & & & &7\\
289&J1805-0619& &10& & & & & & &10\\
290&J1806-2125&19& & & & & & & &19\\
291&J1807-0847& & &21& & & & & &21\\
292&J1808-0813& &12& & & & & & &12\\
293&J1808-3249& &13& & & & & & &13\\
294&J1809-0743& &10& & & & & & &10\\
295&J1811-0154& &12& & & & & & &12\\
296&J1812-1910&13&5& & & & & & &9\\
297&J1814-1744& &8& & & & & & &8\\
298&J1815-1738&34& & & & & & & &34\\
299&J1816-5643& &14& & & & & & &14\\
300&J1817-3837& &25& & & & & & &25\\
301&J1819+1305& &6& & & & & & &6\\
302&J1820-0427& & &33& & & & &41&37\\
303&J1820-1529&24& & & & & & & &24\\
304&J1820-1818& &9& & & & & & &9\\
305&J1821-1419&8& & & & & & & &8\\
306&J1822-2256& &10& & & & & & &10\\
307&J1824-1945&38&53&80& & & & & &57\\
308&J1825-0935& & &21& & &11& & &16\\
309&J1825-1446&26& & &8& & & & &17\\
310&J1826-1334&3&3& &13&14& & & &8\\
311&J1828-1057&7&6& & & & & & &7\\
312&J1829-1751& & &18& & & & & &18\\
313&J1830-1059&38&34& & & & & & &36\\
314&J1832-0827&18&17& & & & & & &18\\
315&J1834-0731&14& & & & & & & &14\\
316&J1835-0643&13& & & & & & & &13\\
317&J1835-0944&14&9& & & & & & &11\\
318&J1835-1106&20&15& &52&73& & & &40\\
319&J1837-0045& &15& & & & & & &15\\
320&J1837-0559&18&9& & & & & & &14\\
321&J1837+1221& &19& & & & & & &19\\
322&J1837-1837& &20& & & & & & &20\\
323&J1838-0453&20&11& & & & & & &15\\
324&J1838-0549&12&13& & & & & & &13\\
325&J1839-0905&17&11& & & & & & &14\\
326&J1841-0345& &8& & & & & & &8\\
327&J1841-0425&20&19& & &41& & & &27\\
328&J1841-7845& &6& & & & & & &6\\
329&J1842-0905&12&11& & & & & & &12\\
330&J1842+1332& &2& & & & & & &2\\
331&J1843-0355&7& & & & & & & &7\\
332&J1843-0702& &22& & & & & & &22\\
333&J1844-0538&20&12& & &27& & & &20\\
334&J1845-0316& &6& & & & & & &6\\
335&J1845-0434&16&16& &29& & & & &20\\
336&J1845-0743&11&10& & & & & & &10\\
337&J1847-0402&14&12& & & & & & &13\\
338&J1848-0123& & &17& & & & & &17\\
339&J1848-1414& &9& & & & & & &9\\
340&J1848-1952& & &25& & &53& &35&37\\
341&J1852-2610& &8& & & & & & &8\\
342&J1853+0011& &19& & & & & & &19\\
343&J1855-0941& &5& & & & & & &5\\
344&J1900-2600& & &9& & & & & &9\\
345&J1901+0331& & &12& & &27&81&66&47\\
346&J1901-0906& &14& & & & & & &14\\
347&J1901-1740& &9& & & & & & &9\\
348&J1903+0135& & &38& & & & &47&42\\
349&J1904+0004& &8& & & & & & &8\\
350&J1904-1224& &18& & & & & & &18\\
351&J1913-0440& & &48& & & & & &48\\
352&J1919+0134& &9& & & & & & &9\\
353&J1932+1059& & &22& & &16& & &19\\
354&J1932-3655& &16& & & & & & &16\\
355&J1941-2602& & &48& & & & & &48\\
356&J1943+0609& &13& & & & & & &13\\
357&J1943-1237& & &39& & & & & &39\\
358&J1944-1750& &5& & & & & & &5\\
359&J1946-1312& &15& & & & & & &15\\
360&J1946+1805& & &11& & &5& & &8\\
361&J1946-2913& &17& & & & & & &17\\
362&J1947+0915& &12& & & & & & &12\\
363&J1949-2524& &38& & & & & & &38\\
364&J1956+0838& &8& & & & & & &8\\
365&J2006-0807& &3& & & & & & &3\\
366&J2007+0809& &1& & & & & & &1\\
367&J2046-0421& & &34& & & & & &34\\
368&J2048-1616& &10&16& & &44& &80&38\\
369&J2053-7200& & &11& & &29& &71&37\\
370&J2108-3429& &26& & & & & & &26\\
371&J2116+1414& &11& & & & & & &11\\
372&J2155-3118& & &25& & &66&73&65&57\\
373&J2248-0101& &18& & & & & & &18\\
374&J2324-6054& & &24& & &65&90&85&66\\
375&J2330-2005& & &33& & & &84&79&65\\
376&J2346-0609& &10& & & & & & &10\\[1mm]
\end{longtable}

\section{Calculations of magnetic inductions}

Now we know values of $\beta$ and can use the formula (3) to calculate magnetic inductions $B_s$ at the surface of the considered pulsars:

\begin{equation}
B_{calc}=\frac{3.2\times 10^{19}\sqrt{P\dot{P}}}{\sin\beta}.			
\end{equation}

\vspace{\baselineskip}
The results are given in the Table 2. The seventh column contains equatorial magnetic inductions from the catalog (Manchester et al., 2005) calculated using $\sin\beta =1$. As we can see our values several times higher than those from the catalog. This effect has been expected, because $\sin\beta <1$ always. Fig.3 shows the distribution of our calculated inductions. For the comparison we put in this figure the similar distribution for the values from the catalog (Manchester et al., 2005). The ratios of these quantities are given in the last column of the Table 2, the mean ratio is $B_{calc}/B_s =5$.

\begin{figure}
\begin{center}
		\includegraphics[width=12cm]{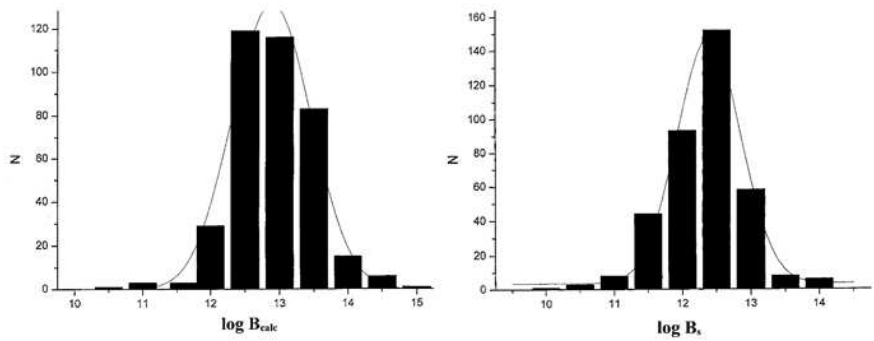}
    \caption{Distributions of calculated equatorial magnetic inductions (left histogram)
and equatorial inductions from the catalog (Manchester et al., 2005) (right).}
    \label{fig:Fig3.png}
		\end{center}
\end{figure}

The analysis of distributions in Fig.3 shows that they can be described well by the following approximations:

\begin{equation}
N\left(\log B_{calc}\right)=132 \exp{ \left\{ - \left[ \frac{\log B_{calc}-12.89}{1.13} \right]^2 \right\}}.			
\end{equation}
\begin{equation}
N\left(\log B_s\right)=144 \exp{ \left\{ - \left[ \frac{\log B_s-12.39}{0.92} \right]^2 \right\}}.			
\end{equation}

\begin{longtable}{|c|c|c|c|c|c|c|c|}
\caption[The magnetic inductions of pulsars.]{The magnetic inductions of pulsars.}
\\
\hline
&$PSR$&$P(s)$&$\dot{P}$&$\left\langle\beta\right\rangle (deg)$&$B_{calc} (G)$&$B_s (G)$&$B_{calc}/B_s$\\
	\hline \endhead
\hline
\endfoot
1&J0034-0721&0.942951&4.08E-16&7&5.04E+12&6.28E+11&8.0\\
2&J0051+0423&0.354732&7.00E-18&5&5.27E+11&5.04E+10&10.4\\
3&J0108-1431&0.807565&7.70E-17&9&1.61E+12&2.52E+11&6.4\\
4&J0134-2937&0.136962&7.84E-17&16&3.74E+11&1.05E+11&3.6\\
5&J0151-0635&1.464665&4.43E-16&4&1.11E+13&8.15E+11&13.6\\
6&J0152-1637&0.832742&1.30E-15&32&1.99E+12&1.05E+12&1.9\\
7&J0206-4028&0.630550&1.20E-15&35&1.52E+12&8.80E+11&1.7\\
8&J0211-8159&1.077333&2.90E-16&5&6.88E+12&5.66E+11&12.2\\
9&J0255-5304&0.447708&3.06E-17&35&2.05E+11&1.18E+11&1.7\\
10&J0304+1932&1.387584&1.30E-15&11&7.39E+12&1.36E+12&5.4\\
11&J0401-7608&0.545253&1.54E-15&23&2.38E+12&9.27E+11&2.6\\
12&J0448-2749&0.450448&1.48E-16&13&1.14E+12&2.62E+11&4.4\\
13&J0450-1248&0.438014&1.03E-16&7&1.68E+12&2.15E+11&7.8\\
14&J0452-1759&0.548939&5.75E-15&15&7.06E+12&1.80E+12&3.9\\
15&J0459-0210&1.133076&1.40E-15&14&5.24E+12&1.27E+12&4.1\\
16&J0520-2553&0.241642&3.01E-17&13&3.80E+11&8.63E+10&4.4\\
17&J0525+1115&0.354438&7.36E-17&13&7.10E+11&1.63E+11&4.4\\
18&J0533+0402&0.963018&1.60E-16&18&1.26E+12&3.97E+11&3.2\\
19&J0536-7543&1.245856&5.65E-16&23&2.16E+12&8.49E+11&2.5\\
20&J0540-7125&1.286015&8.20E-16&11&5.58E+12&1.04E+12&5.4\\
21&J0543+2329&0.245975&1.54E-14&11&1.04E+13&1.97E+12&5.3\\
22&J0601-0527&0.395969&1.30E-15&10&4.01E+12&7.27E+11&5.5\\
23&J0614+2229&0.334960&5.94E-14&26&1.02E+13&4.52E+12&2.3\\
24&J0624-0424&1.039076&8.30E-16&8&6.60E+12&9.40E+11&7.0\\
25&J0630-2834&1.244419&7.12E-15&20&8.89E+12&3.01E+12&3.0\\
26&J0631+1036&0.287800&1.05E-13&10&3.18E+13&5.55E+12&5.7\\
27&J0636-4549&1.984597&3.17E-15&27&5.63E+12&2.54E+12&2.2\\
28&J0656-2228&1.224754&2.67E-17&20&5.32E+11&1.83E+11&2.9\\
29&J0656-5449&0.183157&3.19E-17&14&3.25E+11&7.74E+10&4.2\\
30&J0659+1414&0.384891&5.50E-14&13&2.05E+13&4.66E+12&4.4\\
31&J0709-5923&0.485268&1.26E-16&30&5.00E+11&2.50E+11&2.0\\
32&J0719-2545&0.974725&7.28E-15&16&9.62E+12&2.70E+12&3.6\\
33&J0729-1448&0.251659&1.13E-13&10&3.25E+13&5.40E+12&6.0\\
34&J0729-1836&0.510160&1.90E-14&12&1.57E+13&3.15E+12&5.0\\
35&J0738-4042&0.374920&1.62E-15&16&2.83E+12&7.88E+11&3.6\\
36&J0742-2822&0.166762&1.68E-14&32&3.23E+12&1.69E+12&1.9\\
37&J0745-5353&0.214836&2.73E-15&27&1.71E+12&7.75E+11&2.2\\
38&J0749-4247&1.095452&9.77E-16&20&2.99E+12&1.05E+12&2.9\\
39&J0809-4753&0.547199&3.08E-15&24&3.17E+12&1.31E+12&2.4\\
40&J0820-1350&1.238130&2.11E-15&59&1.92E+12&1.63E+12&1.2\\
41&J0820-3921&1.073567&1.22E-14&4&5.26E+13&3.66E+12&14.4\\
42&J0821-4221&0.396728&3.48E-15&10&6.79E+12&1.19E+12&5.7\\
43&J0834-4159&0.121116&4.44E-15&15&2.96E+12&7.42E+11&4.0\\
44&J0837+0610&1.273768&6.80E-15&32&5.66E+12&2.98E+12&1.9\\
45&J0837-4135&0.751624&3.54E-15&54&2.04E+12&1.65E+12&1.2\\
46&J0838-2621&0.308581&3.90E-17&7&9.32E+11&1.11E+11&8.4\\
47&J0843-5022&0.208956&1.72E-16&14&7.78E+11&1.92E+11&4.1\\
48&J0846-3533&1.116097&1.60E-15&12&6.44E+12&1.35E+12&4.8\\
49&J0849-6322&0.367953&7.91E-16&10&3.09E+12&5.46E+11&5.7\\
50&J0855-3331&1.267536&6.32E-15&47&3.89E+12&2.86E+12&1.4\\
51&J0856-6137&0.962509&1.68E-15&13&5.59E+12&1.29E+12&4.3\\
52&J0857-4424&0.326774&2.33E-14&14&1.19E+13&2.79E+12&4.3\\
53&J0901-4624&0.441995&8.75E-14&11&3.44E+13&6.29E+12&5.5\\
54&J0902-6325&0.660313&1.07E-16&12&1.26E+12&2.69E+11&4.7\\
55&J0905-4536&0.988281&1.49E-16&2&1.06E+13&3.88E+11&27.3\\
56&J0905-5127&0.346287&2.49E-14&20&8.77E+12&2.97E+12&3.0\\
57&J0907-5157&0.253556&1.83E-15&22&1.84E+12&6.89E+11&2.7\\
58&J0908-4913&0.106755&1.52E-14&44&1.85E+12&1.29E+12&1.4\\
59&J0922+0638&0.430627&1.37E-14&42&3.70E+12&2.46E+12&1.5\\
60&J0924-5302&0.746295&3.55E-14&18&1.71E+13&5.21E+12&3.3\\
61&J0924-5814&0.739501&4.88E-15&5&2.33E+13&1.92E+12&12.2\\
62&J0932-3217&1.931627&2.50E-16&30&1.42E+12&7.03E+11&2.0\\
63&J0934-4154&0.570409&2.69E-16&13&1.75E+12&3.96E+11&4.4\\
64&J0934-5249&1.444773&4.65E-15&40&4.08E+12&2.62E+12&1.6\\
65&J0941-5244&0.658558&1.14E-15&15&3.50E+12&8.76E+11&4.0\\
66&J0942-5552&0.664367&2.29E-14&37&6.57E+12&3.94E+12&1.7\\
67&J0942-5657&0.808127&3.96E-14&34&1.04E+13&5.73E+12&1.8\\
68&J0953+0755&0.253065&2.30E-16&15&9.25E+11&2.44E+11&3.8\\
69&J0954-5430&0.472834&4.39E-14&20&1.37E+13&4.61E+12&3.0\\
70&J0955-5304&0.862118&3.52E-15&25&4.21E+12&1.76E+12&2.4\\
71&J1001-5507&1.436583&5.14E-14&47&1.20E+13&8.70E+12&1.4\\
72&J1003-4747&0.307072&2.21E-14&12&1.25E+13&2.63E+12&4.8\\
73&J1015-5719&0.139882&5.74E-14&5&3.52E+13&2.87E+12&12.3\\
74&J1016-5345&0.769584&1.93E-15&29&2.54E+12&1.23E+12&2.1\\
75&J1016-5857&0.107386&8.08E-14&8&2.18E+13&2.98E+12&7.3\\
76&J1017-5621&0.503459&3.13E-15&22&3.38E+12&1.27E+12&2.7\\
77&J1019-5749&0.162499&2.01E-14&7&1.47E+13&1.83E+12&8.1\\
78&J1020-6026&0.140480&6.74E-15&5&1.24E+13&9.85E+11&12.6\\
79&J1032-5911&0.464208&3.00E-15&8&8.53E+12&1.19E+12&7.2\\
80&J1034-3224&1.150590&2.30E-16&2&1.63E+13&5.21E+11&31.2\\
81&J1036-4926&0.510368&1.65E-15&18&3.08E+12&9.28E+11&3.3\\
82&J1038-5831&0.661992&1.25E-15&17&3.12E+12&9.21E+11&3.4\\
83&J1043-6116&0.288602&1.04E-14&19&5.51E+12&1.75E+12&3.1\\
84&J1046-5813&0.369427&1.14E-15&19&2.00E+12&6.58E+11&3.0\\
85&J1047-3032&0.330328&6.10E-17&5&1.68E+12&1.44E+11&11.6\\
86&J1047-6709&0.198451&1.69E-15&14&2.41E+12&5.86E+11&4.1\\
87&J1048-5832&0.123671&9.63E-14&42&5.20E+12&3.49E+12&1.5\\
88&J1052-5954&0.180592&2.00E-14&20&5.63E+12&1.92E+12&2.9\\
89&J1055-6028&0.099661&2.95E-14&12&8.16E+12&1.74E+12&4.7\\
90&J1056-6258&0.422447&3.58E-15&23&3.22E+12&1.24E+12&2.6\\
91&J1057-5226&0.197108&5.83E-15&15&4.15E+12&1.09E+12&3.8\\
92&J1059-5742&1.184999&4.30E-15&54&2.83E+12&2.28E+12&1.2\\
93&J1110-5637&0.558253&2.06E-15&12&5.18E+12&1.09E+12&4.7\\
94&J1112-6613&0.334213&8.24E-16&11&2.89E+12&5.31E+11&5.4\\
95&J1112-6926&0.820484&2.84E-15&12&7.67E+12&1.54E+12&5.0\\
96&J1114-6100&0.880820&4.61E-14&9&4.17E+13&6.45E+12&6.5\\
97&J1115-6052&0.259777&7.23E-15&27&3.08E+12&1.39E+12&2.2\\
98&J1116-4122&0.943158&7.95E-15&46&3.88E+12&2.77E+12&1.4\\
99&J1119-6127&0.407963&4.02E-12&12&2.01E+14&4.10E+13&4.9\\
100&J1123-4844&0.244838&6.54E-17&13&5.63E+11&1.28E+11&4.4\\
101&J1123-6259&0.271434&5.25E-15&26&2.77E+12&1.21E+12&2.3\\
102&J1126-6054&0.202737&2.81E-16&19&7.54E+11&2.42E+11&3.1\\
103&J1126-6942&0.579416&3.30E-15&16&5.22E+12&1.40E+12&3.7\\
104&J1133-6250&1.022875&4.52E-16&2&2.48E+13&6.88E+11&36.1\\
105&J1136+1551&1.187913&3.73E-15&43&3.12E+12&2.13E+12&1.5\\
106&J1137-6700&0.556216&7.17E-17&2&4.64E+12&2.02E+11&23.0\\
107&J1138-6207&0.117564&1.25E-14&30&2.47E+12&1.23E+12&2.0\\
108&J1141-3107&0.538432&1.96E-15&11&5.43E+12&1.04E+12&5.2\\
109&J1141-3322&0.291468&4.63E-16&9&2.28E+12&3.72E+11&6.1\\
110&J1143-5158&0.675646&6.95E-16&22&1.82E+12&6.94E+11&2.6\\
111&J1146-6030&0.273372&1.79E-15&14&3.02E+12&7.09E+11&4.3\\
112&J1157-6224&0.400522&3.93E-15&19&3.87E+12&1.27E+12&3.0\\
113&J1204-6843&0.308861&2.17E-16&18&8.38E+11&2.62E+11&3.2\\
114&J1215-5328&0.636414&1.15E-16&7&2.21E+12&2.74E+11&8.0\\
115&J1216-6223&0.374047&1.68E-14&10&1.50E+13&2.54E+12&5.9\\
116&J1224-6407&0.216476&4.95E-15&26&2.38E+12&1.05E+12&2.3\\
117&J1225-5556&1.018453&2.06E-15&17&4.96E+12&1.47E+12&3.4\\
118&J1225-6408&0.419618&9.46E-16&12&3.15E+12&6.38E+11&4.9\\
119&J1231-4609&0.877239&3.80E-17&7&1.42E+12&1.85E+11&7.7\\
120&J1236-5033&0.294760&1.56E-16&9&1.39E+12&2.17E+11&6.4\\
121&J1240-4124&0.512242&1.74E-15&31&1.83E+12&9.55E+11&1.9\\
122&J1243-6423&0.388481&4.50E-15&59&1.57E+12&1.34E+12&1.2\\
123&J1244-5053&0.275207&1.00E-15&18&1.68E+12&5.31E+11&3.2\\
124&J1248-6344&0.198335&1.69E-14&6&1.68E+13&1.85E+12&9.1\\
125&J1253-5820&0.255496&2.10E-15&15&2.82E+12&7.41E+11&3.8\\
126&J1301-6305&0.184528&2.67E-13&4&1.12E+14&7.10E+12&15.7\\
127&J1305-6203&0.427762&3.21E-14&19&1.16E+13&3.75E+12&3.1\\
128&J1305-6455&0.571647&4.03E-15&6&1.38E+13&1.54E+12&8.9\\
129&J1306-6617&0.473026&5.97E-15&5&1.86E+13&1.70E+12&10.9\\
130&J1319-6056&0.284351&1.53E-15&17&2.31E+12&6.67E+11&3.5\\
131&J1320-3512&0.458488&1.90E-18&10&1.81E+11&2.99E+10&6.0\\
132&J1320-5359&0.279729&9.25E-15&29&3.33E+12&1.63E+12&2.0\\
133&J1326-5859&0.477991&3.24E-15&35&2.22E+12&1.26E+12&1.8\\
134&J1327-6222&0.529913&1.89E-14&20&9.58E+12&3.20E+12&3.0\\
135&J1327-6301&0.196479&1.53E-15&6&4.92E+12&5.55E+11&8.9\\
136&J1327-6400&0.280678&3.12E-14&10&1.70E+13&2.99E+12&5.7\\
137&J1333-4449&0.345603&5.40E-19&24&3.39E+10&1.38E+10&2.5\\
138&J1339-4712&0.137055&5.30E-19&35&1.52E+10&8.62E+09&1.8\\
139&J1340-6456&0.378622&5.05E-15&10&8.18E+12&1.40E+12&5.8\\
140&J1341-6220&0.193340&2.53E-13&47&9.63E+12&7.08E+12&1.4\\
141&J1349-6130&0.259363&5.13E-15&36&1.97E+12&1.17E+12&1.7\\
142&J1352-6803&0.628903&1.23E-15&9&5.96E+12&8.91E+11&6.7\\
143&J1356-5521&0.507380&7.24E-16&9&3.87E+12&6.13E+11&6.3\\
144&J1357-6429&0.166108&3.60E-13&8&5.83E+13&7.83E+12&7.4\\
145&J1359-6038&0.127501&6.34E-15&37&1.50E+12&9.10E+11&1.6\\
146&J1401-6357&0.842790&1.67E-14&35&6.62E+12&3.80E+12&1.7\\
147&J1403-7646&1.306198&1.20E-15&4&1.74E+13&1.27E+12&13.7\\
148&J1406-6121&0.213075&5.47E-14&43&5.03E+12&3.45E+12&1.5\\
149&J1410-7404&0.278729&6.74E-18&88&4.39E+10&4.39E+10&1.0\\
150&J1412-6145&0.315225&9.87E-14&14&2.27E+13&5.64E+12&4.0\\
151&J1413-6141&0.285625&3.33E-13&44&1.42E+13&9.88E+12&1.4\\
152&J1413-6307&0.394946&7.43E-15&33&3.16E+12&1.73E+12&1.8\\
153&J1415-6621&0.392479&5.80E-16&18&1.58E+12&4.83E+11&3.3\\
154&J1427-4158&0.586486&6.21E-16&14&2.61E+12&6.11E+11&4.3\\
155&J1428-5530&0.570290&2.08E-15&31&2.11E+12&1.10E+12&1.9\\
156&J1430-6623&0.785441&2.77E-15&25&3.48E+12&1.49E+12&2.3\\
157&J1452-5851&0.386625&5.07E-14&20&1.31E+13&4.48E+12&2.9\\
158&J1453-6413&0.179485&2.75E-15&40&1.10E+12&7.10E+11&1.5\\
159&J1456-6843&0.263377&9.83E-17&23&4.16E+11&1.63E+11&2.6\\
160&J1507-4352&0.286757&1.60E-15&24&1.72E+12&6.86E+11&2.5\\
161&J1507-6640&0.355655&1.16E-15&43&9.47E+11&6.49E+11&1.5\\
162&J1512-5759&0.128694&6.85E-15&30&1.90E+12&9.50E+11&2.0\\
163&J1513-5908&0.151251&1.53E-12&6&1.53E+14&1.54E+13&9.9\\
164&J1514-4834&0.454839&9.25E-16&21&1.81E+12&6.56E+11&2.8\\
165&J1514-5925&0.148796&2.88E-15&14&2.79E+12&6.63E+11&4.2\\
166&J1515-5720&0.286646&6.10E-15&13&5.93E+12&1.34E+12&4.4\\
167&J1517-4356&0.650837&2.16E-16&18&1.24E+12&3.79E+11&3.3\\
168&J1522-5829&0.395353&2.00E-15&9&5.47E+12&9.01E+11&6.1\\
169&J1524-5706&1.116049&3.56E-13&15&7.96E+13&2.02E+13&3.9\\
170&J1528-4109&0.526556&3.96E-16&21&1.31E+12&4.62E+11&2.8\\
171&J1530-5327&0.278957&4.68E-15&17&4.02E+12&1.16E+12&3.5\\
172&J1531-4012&0.356849&9.63E-17&16&6.63E+11&1.88E+11&3.5\\
173&J1534-5334&1.368881&1.43E-15&19&4.25E+12&1.41E+12&3.0\\
174&J1534-5405&0.289689&1.54E-15&10&3.97E+12&6.77E+11&5.9\\
175&J1535-4114&0.432866&4.07E-15&12&6.37E+12&1.34E+12&4.8\\
176&J1536-3602&1.319759&7.90E-16&6&9.66E+12&1.03E+12&9.4\\
177&J1538-5551&0.104675&3.21E-15&68&6.35E+11&5.86E+11&1.1\\
178&J1539-5626&0.243392&4.85E-15&20&3.15E+12&1.10E+12&2.9\\
179&J1541-5535&0.295838&7.50E-14&26&1.09E+13&4.77E+12&2.3\\
180&J1542-5034&0.599245&3.97E-15&33&2.89E+12&1.56E+12&1.9\\
181&J1543-5459&0.377119&5.20E-14&46&6.26E+12&4.48E+12&1.4\\
182&J1548-5607&0.170934&1.07E-14&9&8.57E+12&1.37E+12&6.3\\
183&J1549-4848&0.288347&1.41E-14&17&7.03E+12&2.04E+12&3.4\\
184&J1551-5310&0.453394&1.95E-13&19&2.95E+13&9.52E+12&3.1\\
185&J1557-4258&0.329187&3.30E-16&13&1.54E+12&3.34E+11&4.6\\
186&J1559-4438&0.257056&1.02E-15&25&1.24E+12&5.18E+11&2.4\\
187&J1600-5044&0.192601&5.06E-15&22&2.64E+12&9.99E+11&2.6\\
188&J1600-5751&0.194454&2.13E-15&5&6.92E+12&6.51E+11&10.6\\
189&J1601-5335&0.288457&6.24E-14&13&1.89E+13&4.29E+12&4.4\\
190&J1602-5100&0.864227&6.96E-14&32&1.49E+13&7.85E+12&1.9\\
191&J1603-3539&0.141909&1.24E-16&8&9.80E+11&1.34E+11&7.3\\
192&J1603-5657&0.496077&2.79E-15&54&1.47E+12&1.19E+12&1.2\\
193&J1604-4909&0.327418&1.02E-15&31&1.15E+12&5.85E+11&2.0\\
194&J1605-5257&0.658013&2.56E-16&5&5.29E+12&4.15E+11&12.8\\
195&J1607-0032&0.421816&3.06E-16&26&8.17E+11&3.64E+11&2.2\\
196&J1609-1930&1.557917&5.09E-16&28&1.95E+12&9.01E+11&2.2\\
197&J1611-5209&0.182492&5.17E-15&60&1.14E+12&9.83E+11&1.2\\
198&J1612-2408&0.923834&1.57E-15&15&4.83E+12&1.22E+12&4.0\\
199&J1614-3937&0.407292&1.59E-16&11&1.31E+12&2.57E+11&5.1\\
200&J1614-5048&0.231694&4.95E-13&44&1.55E+13&1.08E+13&1.4\\
201&J1615-5537&0.791526&2.00E-15&18&4.09E+12&1.27E+12&3.2\\
202&J1626-4807&0.293928&1.75E-14&8&1.74E+13&2.29E+12&7.6\\
203&J1632-4757&0.228564&1.51E-14&16&6.95E+12&1.88E+12&3.7\\
204&J1632-4818&0.813453&6.50E-13&17&8.07E+13&2.33E+13&3.5\\
205&J1633-5015&0.352142&3.79E-15&14&4.87E+12&1.17E+12&4.2\\
206&J1637-4553&0.118771&3.19E-15&18&1.98E+12&6.23E+11&3.2\\
207&J1637-4642&0.154027&5.92E-14&12&1.47E+13&3.06E+12&4.8\\
208&J1638-4417&0.117802&1.61E-15&12&2.17E+12&4.40E+11&4.9\\
209&J1638-4608&0.278137&5.15E-14&66&4.20E+12&3.83E+12&1.1\\
210&J1638-4725&0.763933&4.79E-15&9&1.22E+13&1.93E+12&6.3\\
211&J1639-4604&0.529116&5.78E-15&10&9.84E+12&1.77E+12&5.6\\
212&J1640-4715&0.517405&4.20E-14&15&1.86E+13&4.72E+12&3.9\\
213&J1641-2347&1.091008&4.11E-17&7&1.64E+12&2.14E+11&7.7\\
214&J1643-4505&0.237383&3.18E-14&15&1.09E+13&2.78E+12&3.9\\
215&J1644-4559&0.455060&2.01E-14&7&2.68E+13&3.06E+12&8.8\\
216&J1645-0317&0.387690&1.78E-15&8&6.32E+12&8.41E+11&7.5\\
217&J1646-4346&0.231603&1.13E-13&17&1.75E+13&5.17E+12&3.4\\
218&J1646-6831&1.785611&1.70E-15&54&2.18E+12&1.76E+12&1.2\\
219&J1648-4611&0.164950&2.37E-14&16&7.41E+12&2.00E+12&3.7\\
220&J1649-4653&0.557019&4.97E-14&20&1.57E+13&5.33E+12&2.9\\
221&J1649-5553&0.613571&1.70E-15&3&2.03E+13&1.03E+12&19.7\\
222&J1650-1654&1.749552&3.20E-15&12&1.19E+13&2.40E+12&5.0\\
223&J1650-4502&0.380870&1.61E-14&48&3.39E+12&2.50E+12&1.4\\
224&J1650-4921&0.156399&1.82E-15&33&9.79E+11&5.40E+11&1.8\\
225&J1651-4246&0.844081&4.81E-15&9&1.34E+13&2.04E+12&6.6\\
226&J1651-7642&1.755311&1.36E-15&7&1.28E+13&1.57E+12&8.1\\
227&J1652-1400&0.305447&1.76E-17&10&4.47E+11&7.42E+10&6.0\\
228&J1653-3838&0.305037&2.79E-15&16&3.29E+12&9.33E+11&3.5\\
229&J1654-2713&0.791822&1.68E-16&16&1.35E+12&3.69E+11&3.7\\
230&J1655-3048&0.542936&3.66E-17&3&2.86E+12&1.43E+11&20.0\\
231&J1700-3312&1.358307&4.71E-15&12&1.25E+13&2.56E+12&4.9\\
232&J1701-3726&2.454609&1.11E-14&9&3.28E+13&5.29E+12&6.2\\
233&J1701-4533&0.322909&5.19E-16&7&3.47E+12&4.14E+11&8.4\\
234&J1702-4128&0.182136&5.23E-14&21&8.59E+12&3.12E+12&2.8\\
235&J1702-4306&0.215507&9.79E-15&11&7.39E+12&1.47E+12&5.0\\
236&J1702-4310&0.240524&2.24E-13&11&4.06E+13&7.42E+12&5.5\\
237&J1703-3241&1.211785&6.60E-16&53&1.13E+12&9.05E+11&1.3\\
238&J1703-4851&1.396401&5.08E-15&10&1.53E+13&2.70E+12&5.7\\
239&J1705-1906&0.298987&4.14E-15&41&1.71E+12&1.13E+12&1.5\\
240&J1705-3950&0.318941&6.06E-14&22&1.18E+13&4.45E+12&2.6\\
241&J1709-1640&0.653054&6.31E-15&58&2.42E+12&2.05E+12&1.2\\
242&J1709-4429&0.102459&9.30E-14&16&1.14E+13&3.12E+12&3.6\\
243&J1714-1054&0.696279&5.88E-17&27&4.59E+11&2.05E+11&2.2\\
244&J1715-3903&0.278481&3.77E-14&12&1.63E+13&3.28E+12&5.0\\
245&J1717-5800&0.321793&1.96E-16&6&2.52E+12&2.54E+11&9.9\\
246&J1718-3718&3.378574&1.61E-12&18&2.48E+14&7.47E+13&3.3\\
247&J1719-4006&0.189094&1.67E-15&14&2.43E+12&5.68E+11&4.3\\
248&J1721-3532&0.280424&2.52E-14&24&6.58E+12&2.69E+12&2.4\\
249&J1722-3207&0.477158&6.46E-16&29&1.14E+12&5.62E+11&2.0\\
250&J1722-3632&0.399183&4.46E-15&7&1.08E+13&1.35E+12&8.0\\
251&J1722-3712&0.236173&1.09E-14&37&2.73E+12&1.62E+12&1.7\\
252&J1723-3659&0.202722&8.01E-15&20&3.73E+12&1.29E+12&2.9\\
253&J1726-3530&1.110132&1.22E-12&40&5.77E+13&3.72E+13&1.6\\
254&J1730-3350&0.139460&8.48E-14&44&5.01E+12&3.48E+12&1.4\\
255&J1731-4744&0.829829&1.64E-13&28&2.48E+13&1.18E+13&2.1\\
256&J1733-2228&0.871683&4.27E-17&5&2.20E+12&1.95E+11&11.3\\
257&J1733-3716&0.337586&1.50E-14&12&1.13E+13&2.28E+12&5.0\\
258&J1734-3333&1.169341&2.28E-12&13&2.26E+14&5.22E+13&4.3\\
259&J1735-3258&0.350963&2.61E-14&10&1.69E+13&3.06E+12&5.5\\
260&J1737-3137&0.450432&1.39E-13&15&3.09E+13&8.00E+12&3.9\\
261&J1737-3555&0.397585&6.12E-15&19&4.77E+12&1.58E+12&3.0\\
262&J1738-2955&0.443398&8.19E-14&21&1.74E+13&6.10E+12&2.8\\
263&J1739+0612&0.234169&1.56E-16&10&1.08E+12&1.94E+11&5.6\\
264&J1739-1313&1.215698&8.17E-17&50&4.15E+11&3.19E+11&1.3\\
265&J1739-2903&0.322882&7.88E-15&23&4.06E+12&1.61E+12&2.5\\
266&J1739-3023&0.114368&1.14E-14&23&3.00E+12&1.16E+12&2.6\\
267&J1740-3015&0.606887&4.66E-13&42&2.56E+13&1.70E+13&1.5\\
268&J1741-3927&0.512211&1.93E-15&56&1.22E+12&1.01E+12&1.2\\
269&J1742-4616&0.412401&3.38E-17&6&1.13E+12&1.19E+11&9.5\\
270&J1743-3150&2.414576&1.21E-13&13&7.91E+13&1.73E+13&4.6\\
271&J1743-3153&0.193105&1.06E-14&7&1.13E+13&1.45E+12&7.8\\
272&J1743-4212&0.306167&7.83E-16&13&2.15E+12&4.96E+11&4.3\\
273&J1745-3040&0.367429&1.07E-14&19&6.25E+12&2.00E+12&3.1\\
274&J1749-3002&0.609874&7.87E-15&4&3.18E+13&2.22E+12&14.3\\
275&J1750-3157&0.910363&1.97E-16&6&4.43E+12&4.28E+11&10.4\\
276&J1751-4657&0.742352&1.29E-15&34&1.79E+12&9.91E+11&1.8\\
277&J1752-2806&0.562558&8.13E-15&41&3.29E+12&2.16E+12&1.5\\
278&J1755-2534&0.233541&1.12E-14&6&1.64E+13&1.64E+12&10.0\\
279&J1756-2225&0.404980&5.27E-14&12&2.19E+13&4.67E+12&4.7\\
280&J1757-2421&0.234101&1.29E-14&9&1.14E+13&1.76E+12&6.5\\
281&J1759-2302&0.810718&1.07E-14&4&4.37E+13&2.99E+12&14.6\\
282&J1801-2154&0.375297&1.60E-14&24&6.13E+12&2.48E+12&2.5\\
283&J1801-2304&0.415827&1.13E-13&11&3.53E+13&6.93E+12&5.1\\
284&J1801-2451&0.124924&1.28E-13&15&1.61E+13&4.04E+12&4.0\\
285&J1801-2920&1.081908&3.29E-15&9&1.22E+13&1.91E+12&6.4\\
286&J1803-2137&0.133667&1.34E-13&8&3.16E+13&4.29E+12&7.4\\
287&J1803-2712&0.334415&1.71E-17&7&6.26E+11&7.66E+10&8.2\\
288&J1805-0619&0.454651&9.69E-16&10&3.73E+12&6.72E+11&5.6\\
289&J1806-2125&0.481789&1.21E-13&19&2.41E+13&7.74E+12&3.1\\
290&J1807-0847&0.163727&2.88E-17&21&1.96E+11&6.95E+10&2.8\\
291&J1808-0813&0.876044&1.24E-15&12&5.13E+12&1.05E+12&4.9\\
292&J1808-3249&0.364912&7.05E-15&13&7.10E+12&1.62E+12&4.4\\
293&J1809-0743&0.313886&1.52E-16&10&1.28E+12&2.21E+11&5.8\\
294&J1811-0154&0.924945&1.61E-15&12&5.81E+12&1.23E+12&4.7\\
295&J1812-1910&0.430991&3.77E-14&9&2.52E+13&4.08E+12&6.2\\
296&J1814-1744&3.975905&7.45E-13&8&4.07E+14&5.51E+13&7.4\\
297&J1815-1738&0.198436&7.79E-14&34&7.14E+12&3.98E+12&1.8\\
298&J1816-5643&0.217923&1.93E-18&14&8.55E+10&2.08E+10&4.1\\
299&J1817-3837&0.384487&5.80E-16&25&1.14E+12&4.78E+11&2.4\\
300&J1819+1305&1.060364&3.59E-16&6&6.34E+12&6.25E+11&10.1\\
301&J1820-0427&0.598076&6.33E-15&37&3.29E+12&1.97E+12&1.7\\
302&J1820-1529&0.333243&3.79E-14&24&8.83E+12&3.60E+12&2.5\\
303&J1820-1818&0.309905&9.36E-17&9&1.11E+12&1.72E+11&6.5\\
304&J1821-1419&1.656010&8.95E-13&8&2.78E+14&3.89E+13&7.1\\
305&J1822-2256&1.874269&1.35E-15&10&9.48E+12&1.61E+12&5.9\\
306&J1824-1945&0.189335&5.23E-15&57&1.20E+12&1.01E+12&1.2\\
307&J1825-0935&0.769006&5.25E-14&16&2.38E+13&6.43E+12&3.7\\
308&J1825-1446&0.279187&2.27E-14&17&8.76E+12&2.55E+12&3.4\\
309&J1826-1334&0.101487&7.53E-14&8&1.96E+13&2.80E+12&7.0\\
310&J1828-1057&0.246328&2.07E-14&7&1.96E+13&2.29E+12&8.6\\
311&J1829-1751&0.307133&5.55E-15&18&4.25E+12&1.32E+12&3.2\\
312&J1830-1059&0.405043&6.00E-14&36&8.51E+12&4.99E+12&1.7\\
313&J1832-0827&0.647293&6.39E-14&18&2.15E+13&6.51E+12&3.3\\
314&J1834-0731&0.512980&5.82E-14&14&2.25E+13&5.53E+12&4.1\\
315&J1835-0643&0.305830&4.05E-14&13&1.54E+13&3.56E+12&4.3\\
316&J1835-0944&0.145347&4.39E-15&11&4.18E+12&8.08E+11&5.2\\
317&J1835-1106&0.165907&2.06E-14&40&2.91E+12&1.87E+12&1.6\\
318&J1837-0045&0.617037&1.68E-15&15&3.87E+12&1.03E+12&3.8\\
319&J1837-0559&0.201063&3.30E-15&14&3.51E+12&8.25E+11&4.3\\
320&J1837+1221&1.963532&6.20E-15&19&1.11E+13&3.53E+12&3.1\\
321&J1837-1837&0.618358&5.50E-15&20&5.38E+12&1.87E+12&2.9\\
322&J1838-0453&0.380831&1.16E-13&15&2.59E+13&6.72E+12&3.9\\
323&J1838-0549&0.235303&3.34E-14&13&1.28E+13&2.84E+12&4.5\\
324&J1839-0905&0.418969&2.60E-14&14&1.40E+13&3.34E+12&4.2\\
325&J1841-0345&0.204068&5.79E-14&8&2.66E+13&3.48E+12&7.6\\
326&J1841-0425&0.186149&6.39E-15&27&2.45E+12&1.10E+12&2.2\\
327&J1841-7845&0.353603&1.60E-16&6&2.44E+12&2.41E+11&10.1\\
328&J1842-0905&0.344643&1.05E-14&12&9.51E+12&1.92E+12&5.0\\
329&J1842+1332&0.471604&2.29E-16&2&1.05E+13&3.33E+11&31.6\\
330&J1843-0355&0.132314&1.04E-15&7&2.99E+12&3.75E+11&8.0\\
331&J1843-0702&0.191614&2.14E-15&22&1.76E+12&6.48E+11&2.7\\
332&J1844-0538&0.255699&9.71E-15&20&4.68E+12&1.59E+12&2.9\\
333&J1845-0316&0.207636&8.86E-15&6&1.25E+13&1.37E+12&9.1\\
334&J1845-0434&0.486751&1.13E-14&20&6.84E+12&2.38E+12&2.9\\
335&J1845-0743&0.104695&3.67E-16&10&1.11E+12&1.98E+11&5.6\\
336&J1847-0402&0.597769&5.17E-14&13&2.54E+13&5.63E+12&4.5\\
337&J1848-0123&0.659432&5.25E-15&17&6.57E+12&1.88E+12&3.5\\
338&J1848-1414&0.297770&1.41E-17&9&4.26E+11&6.55E+10&6.5\\
339&J1848-1952&4.308190&2.33E-14&37&1.67E+13&1.01E+13&1.7\\
340&J1852-2610&0.336337&8.77E-17&8&1.28E+12&1.74E+11&7.3\\
341&J1853+0011&0.397882&3.35E-14&19&1.11E+13&3.70E+12&3.0\\
342&J1855-0941&0.345401&2.40E-16&5&3.35E+12&2.91E+11&11.5\\
343&J1900-2600&0.612209&2.05E-16&9&2.20E+12&3.58E+11&6.2\\
344&J1901+0331&0.655450&7.46E-15&47&3.08E+12&2.24E+12&1.4\\
345&J1901-0906&1.781928&1.64E-15&14&7.15E+12&1.73E+12&4.1\\
346&J1901-1740&1.956858&8.23E-16&9&8.56E+12&1.28E+12&6.7\\
347&J1903+0135&0.729304&4.03E-15&42&2.57E+12&1.74E+12&1.5\\
348&J1904+0004&0.139525&1.18E-16&8&8.91E+11&1.30E+11&6.9\\
349&J1904-1224&0.750808&7.42E-16&18&2.51E+12&7.55E+11&3.3\\
350&J1913-0440&0.825936&4.07E-15&48&2.51E+12&1.85E+12&1.4\\
351&J1919+0134&1.603984&5.89E-16&9&6.28E+12&9.84E+11&6.4\\
352&J1932+1059&0.226518&1.16E-15&19&1.63E+12&5.18E+11&3.1\\
353&J1932-3655&0.571420&2.84E-16&16&1.47E+12&4.08E+11&3.6\\
354&J1941-2602&0.402858&9.56E-16&46&8.71E+11&6.28E+11&1.4\\
355&J1943+0609&0.446226&4.66E-16&13&2.01E+12&4.61E+11&4.4\\
356&J1943-1237&0.972429&1.66E-15&39&2.04E+12&1.28E+12&1.6\\
357&J1944-1750&0.841158&9.86E-16&5&1.08E+13&9.22E+11&11.7\\
358&J1946-1312&0.491865&1.99E-15&15&3.91E+12&1.00E+12&3.9\\
359&J1946+1805&0.440618&2.41E-17&8&7.38E+11&1.04E+11&7.1\\
360&J1946-2913&0.959448&1.49E-15&17&4.15E+12&1.21E+12&3.4\\
361&J1947+0915&1.480744&4.78E-16&12&4.02E+12&8.51E+11&4.7\\
362&J1949-2524&0.957617&3.27E-15&38&2.93E+12&1.79E+12&1.6\\
363&J1956+0838&0.303911&2.20E-16&8&1.85E+12&2.62E+11&7.1\\
364&J2006-0807&0.580871&4.60E-17&3&2.98E+12&1.65E+11&18.1\\
365&J2007+0809&0.325724&1.37E-16&1&8.90E+12&2.14E+11&41.6\\
366&J2046-0421&1.546938&1.47E-15&34&2.75E+12&1.53E+12&1.8\\
367&J2048-1616&1.961572&1.10E-14&38&7.70E+12&4.69E+12&1.6\\
368&J2053-7200&0.341336&1.96E-16&37&4.32E+11&2.62E+11&1.6\\
369&J2108-3429&1.423102&3.50E-15&26&5.24E+12&2.26E+12&2.3\\
370&J2116+1414&0.440153&2.89E-16&11&1.91E+12&3.61E+11&5.3\\
371&J2155-3118&1.030002&1.24E-15&57&1.36E+12&1.14E+12&1.2\\
372&J2248-0101&0.477233&6.59E-16&18&1.88E+12&5.68E+11&3.3\\
373&J2324-6054&2.347486&2.59E-15&66&2.74E+12&2.49E+12&1.1\\
374&J2330-2005&1.643622&4.63E-15&65&3.07E+12&2.79E+12&1.1\\
375&J2346-0609&1.181463&1.36E-15&10&7.77E+12&1.28E+12&6.1\\ [1mm]
\end{longtable}

\section{Conclusions}
	\begin{enumerate}
		\item Some methods for calculations of the angle $\beta$ between rotation and magnetic axes were applied to obtain the values of $\beta$ for 376 radio pulsars. The distribution of these values shows the predominance of small inclinations of the magnetic axes.
		\item Magnetic inductions at the surface of 375 pulsars considered were calculated. There is no the measured derivative $\dot{P}$ for the pulsar J1713-3949 and it is excluded from the consideration. The distribution of the calculated magnetic inductions can be described by the Gaussian with the maximal value of $10^{13}G$ and the width in the logarithmic scale nearly 1.
		\item The calculated inductions are higher than the catalog equatorial inductions with the mean value of the ratio of these quantities of 5. For the pulsar J1410-7404 $B_{calc} = B_s$. The maximal value of the ratio $B_{calc}/B_s = 41.6$ for the pulsar J2007+0809.
\end{enumerate}

\section{Acknowledgements}

This work has been carried out with the financial support of Basic Research Program of the Presidium of the Russian Academy of Sciences \textbf{\textit{Transitional and Explosive Processes in Astrophysics}} (P-41). We thank A.V.Biryukov for very useful comments and discussions.


\begin{thebibliography}{99}
\bibitem{Keith2010}
Keith M.J., Johnston S., Weltevrede P. and Kramer M., 2010, MNRAS, 402, 745
\bibitem{Kuz'min1983}
Kuz'min A.D., Dagkesamanskaya I.M., 1983, Soviet Astron. Letters, 9, 80
\bibitem{Kuz'min1984}
Kuz'min A.D., Dagkesamanskaya I.M., Pugachev V.D., 1984, Soviet Astron. Letters, 10, 357
\bibitem{Lyne1988}
Lyne A.G., Manchester R.N., 1988, MNRAS, 243, 477
\bibitem{Manchester1977}
Manchester R.N., Taylor J.H., 1977, Pulsars. W.H.Freeman and Company, San Francisco
\bibitem{Manchester2013}
Manchester R.N. et al., 2005, Astron. J., 129, 1993.
\bibitem{Malov2011}
Malov I.F., Nikitina E.B., 2011a, Astron.Rep., 55, 19
\bibitem{Malov2011}
Malov I.F., Nikitina E.B., 2011b, Astron.Rep., 55, 878
\bibitem{Malov2013}
Malov I.F., Nikitina E.B., 2013, Astron.Rep., 57, 833
\bibitem{van Ommen 1977}
van Ommen T.D. et al., 1997, MNRAS, 287, 1210
\bibitem{Weltevrede2008}
Weltevrede P., Johnston S., 2008, MNRAS, 391, 1210
\end{thebibliography}
\end{document}